# Monofiber optical probe using Doppler signals detection for Drop Size and Velocity measurement in air assisted atomization

Mathieu Alonzo*[1], Anthony Lefebvre[2], Zhujun Huang[1], Stéphane Gluck[2], Alain Cartellier[1]
[1] Univ. Grenoble Alpes, CNRS, Grenoble INP**, LEGI, 38000 Grenoble, France
[2] A2 Photonic Sensors, 38016 Grenoble, France

**Abstract**

Reliable measurement of droplet/bubble size and velocity distributions in dense flows is desired in a variety of research fields, both for laboratory and industrial use. A new type of single-mode monofiber optical probe manufactured by *A2 Photonic Sensors* is introduced in this paper: it combines traditional phase detection with the collection of a Doppler signal returned by an incoming gas-liquid interface to provide information on residence times, drop concentration and velocity, which afford then drop chords and liquid flux measurements. Compared with classical optical probes, that new sensor does not require any calibration. The purpose of the present work is to test this technique in assisted atomization in order to provide a mean for spray characterization and ultimately to improve our understanding of atomization mechanisms. The probe has been tested downstream of a coaxial air-assisted atomizer operated at liquid velocity $U_L = 0.14$ to $0.42 \text{ m/s}$, and gas velocity from $U_G = 32$ to about $200 \text{ m/s}$. We first analyzed raw signals in various flow conditions. It happens that, when increasing the gas velocity and the number density of drops, the signal experiences very strong fluctuations of the gas level, making the identification of individual droplets more difficult. That leads us to develop a new signal processing routine specifically adapted to such complex working conditions. At $U_G = 48 \, m/s$ and for three liquid flow rates, the spatial integration of local liquid fluxes represents 92 to 99% of the injected liquid flow rate. These good results demonstrate that the Doppler probe provides reliable statistics on drops velocity and size.

**Introduction**

Assisted atomization consists in destabilizing a liquid jet and breaking it up into droplets by a fast co-axial gas stream, allowing for two-phase mixing at high velocities. The downstream spray formation results from interfacial instabilities [5-7]. Operational and environmental conditions, such as nozzle geometry, gas and liquid flow rate, ambient pressure, turbulence and liquid rheology, etc. have an impact on the droplets size and velocity. Understanding the physical mechanisms by which the liquid jet progressively breaks into lumps of liquid and into droplets is necessary to better design injectors and/or to optimize industrial applications. Droplets characterisation with high accuracy measurements is therefore of great importance. Monofiber optical probes have been widely used for spray measurements in the field of assisted atomization, as they allow to simultaneously measure chord lengths, velocities and *in fine* void fraction and fluxes of droplets [1-3, 9]. The advent of new optical probes that provide Doppler signals for interface velocity measurements requires an experimental validation of the technique. The major benefit of M2 probes is the reliability in velocity measurements since it does not depend on any calibration, together with its ability to measure small drops thanks to its short sensitive length. A recent study proved the reliability of this probe in heterogeneous bubbly flows [4]. In this work, the reliability of this sensor is evaluated with respect to drop detection. First, raw signals delivered by the probe M2 for various gas flow rates at a fixed liquid flow rate are analysed, using a suitable processing routine dedicated to drop detection. Then velocity distributions from Doppler signals are provided. Finally, radial profiles of local variables are measured downstream of the nozzle, under different operating conditions. The performances of this new probe are then asserted by comparing the global liquid flux calculated from the spatial integration of local fluxes with the liquid flow rate injected in the atomizer.

**Material and Methods**

1. **Experimental set-up**

The coaxial injector used in this study includes a 5mm diameter liquid (water) injector surrounded by a $5\ mm$ wide annular gas one (air). Channels are about $1\ m$ long, allowing the flow to fully develop before the exit. A detailed description of this injector has been presented in our previous study [10]. Air is fed by a compressor at a flow rate controlled with a mass flow meter (Brooks Instruments SLA5853S). The gas flow rate evolves from 300 to 2670 *L/min*. Pure water ($18\ M\Omega$ resistivity) is injected by gravity at a liquid flow rate controlled with a gear flow meter (Oval LSF445). It evolves from 10 to 30 L/h, corresponding to a mean liquid velocity at the exit in the range of $0.14 < U_L < 0.42\ m/s$. The optical probe has a conical shape: its latency length is 6 *μm*. In this study, the probe faces the injector and its axis is parallel to the symmetry axis of the injector. The choice of axial distance z between the injector exit and the probe tip is of great importance. Indeed, the length $L_B$ of the intact liquid structure before break-up varies with injection conditions. In order to characterize only the droplets formed by atomization, the probe must be located far enough from the injector to avoid detecting liquid packets that are still connected with the injector. For each injection condition, the time-average breaking length $< L_B >$ was determined by image analysis based on 6000 images acquired at 3000 to 5000 *im/s*. Measurements were collected beyond the break-up length, at a distance z from the injector between 1.5 and 3 times $< L_B >$.

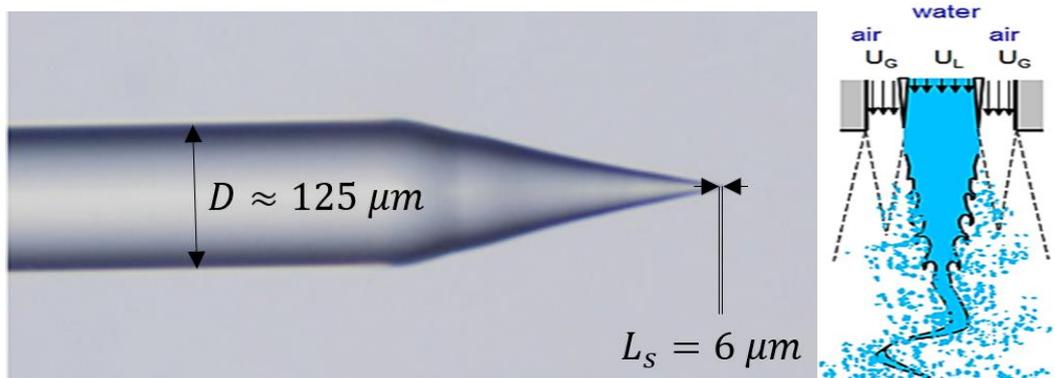

**Figure 1.** View of the probe tip under a microscope. $L_S$ is the latency length, $D$ its diameter (left). Scheme of the co-axial injector (right).

2. **Doppler optical probe: principle and processing routine**

The probe (called M2) is an optical fiber (outer diameter = 125μm) with a conical tip. The fiber is glued inside an inox tube (diameter 2mm). The fiber length exposed to the flow is typically about 4 mm. It is fed with a $\lambda_0 = 1550nm$ wavelength single mode laser light. Some of this light is reflected at the tip. The intensity of the returned signal depends on the refractive index of the medium the probe is located in (see Figure 2). A typical signal recorded when the probe interacts with a drop is shown Figure 2. The probe is initially in the air and the signal amplitude is constant, equal to $V_{G\,local}$. The gas to liquid transition occurs at $t_{entry}$, after which the signal amplitude drops down to the liquid level during $t_L$ i.e. while the probe tips remains within the drop. The signal amplitude increases again up to the gas level at $t_{exit}$, when the sensitive tip leaves the drop. This is the classical response of a phase detection probe. The new feature here is the presence of a Doppler modulation (Figure 2). That signature is due to the use of single mode radiation along with the optimized conical shape of the probe. Beside, this Doppler burst results from the incoming gas to liquid interface, and it is thus recorded just before the gas to liquid transition. Note that a Doppler signal can also arise from the liquid to gas interface, but its amplitude is significantly lower. The Doppler frequency for an incoming plane gas to liquid interface can be expressed as follows:

$$f_D = 2.\frac{V.|\mathbf{n}\cdot\mathbf{k}|}{\lambda_0/n_{ext}} \qquad (1)$$

where V is the interface velocity, **n** the normal to the interface at the contact with the probe tip, **k** the wave number. $\lambda_0$ is the emitted light wavelength and $n_{ext}$ the refractive index of the gaseous phase. This relation may be applied to a drop only when the radius of curvature of the interface is large compared with the probe size. The processing routine for velocity measurement consists basically in extracting Doppler signals presenting a minimum number of successive periods with a good temporal homogeneity. Optimization of this processing routine has been performed and allows measuring velocity dynamics up to 15. Based on measured velocities and liquid residence times, the following local variables are calculated:

- Chord of drop $i$: $C_i = V_i\,T_{Li}$ where $T_{Li}$ is the liquid residence time of drop $i$ and $V_i$ the interface velocity obtained from Doppler analysis.
- Liquid fraction $\alpha_L = \sum T_{Li}/T_t$ where $T_t$ is the acquisition duration.
- The local liquid flux $j_L = \sum C_i/T_t$.

For each acquisition, data convergence is ensured by plotting the evolution of mean values of local variables (size, velocity and liquid residence time). Acquisition durations are set so that data converge toward the mean within 5%. Let us mention in passing that Doppler frequencies are quite high: in the present experimental conditions, the sampling frequency was set at 500 *MHz*.

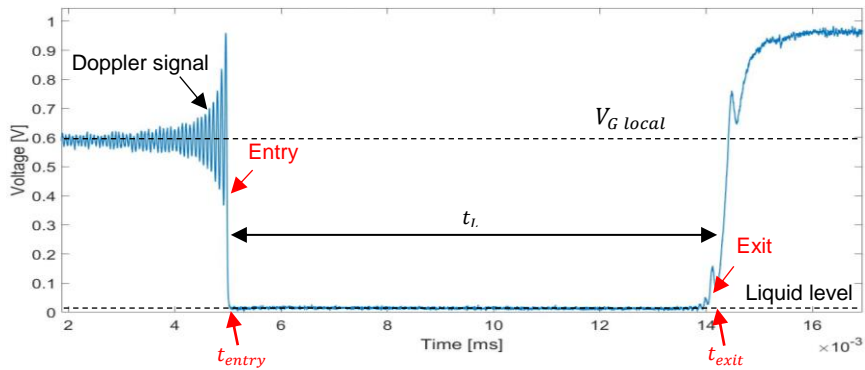

**Figure 2.** Typical signal of a drop going through the probe tip for a time $t_L = t_{exit} - t_{entry}$, corresponding to the time during which the tip is inside the drop, thus fully wetted.

### 1. Evaluation of local and global liquid fluxes

To test the performances of the Doppler probe, our objective is to compare the global liquid flow rate injected in the atomizer with the integral of local liquid fluxes measured along a spray diameter. The first step is to assess the local flux estimation. Indeed, the local flux determination requires accounting for all drops that hit the sensitive tip of the sensor during the measuring duration, the contribution of each drop being the chord cut by the probe extremity. For drops with unambiguous signatures, the output of the signal processing consists in the liquid residence time and the drop velocity, from which the chord is directly deduced. For drops with somewhat distorted signatures, no velocity can be measured and only the liquid residence time is available. For these events, a strategy must be implemented to provide an estimate of the velocity. We have followed the routine proposed by *Hong et al.* (2004) [1]. Chords that are directly measured are correlated with the liquid residence time according to a power law:

$$C_L(i) = a.T_L(i)^b \qquad (2)$$

where the constants *a, b* are fitted. The remaining liquid residence times are transformed into chords using that power law.

The global flux Q can then be computed as the spatial integration of the local fluxes $j_L(r)$ measured along a diameter from $-R_{max}$ to $+R_{max}$:

$$Q = \pi \int_{-R_{max}}^{+R_{max}} j_L(r).r.dr \qquad (3)$$

where *r* is the radial position of the probe from the center of the spray. Polar coordinates are used assuming the spray to be axisymmetric. Discretization of the integral according to the trapezoidal method (Equation 3) allows computing the flux with the assumption that the data are collected over the diameter with a regular spatial step $\Delta R = R_{max}/N$:

$$Q \approx \frac{\pi.R_{max}}{2N} \sum_{i=1}^{2N} \{j_L(R_i).R_i + j_L(R_{i+1}).R_{i+1}\} \qquad (4)$$

where *2N+1* measurement locations from $-R_{max}$ to $+R_{max}$ are considered. Experimental $\Delta R$ was about 1 mm. To minimise errors when performing this integration, one also has to find the spray center where the local flux reaches its maximum. Several acquisitions were performed along two orthogonal directions starting on an approximate position under the injector and the location of the maximum flux was then found. For each run, the distance to the axis was increased until the local flux becomes lower than 10% of the maximum flux at the center of spray: that criterion defines the maximum extent $R_{max}$ of the measuring zone. Beyond this extent, the non-measured flux has a contribution to the global flux below 3%.

2. **Signal characterization for different injection conditions**

The signal exemplified in Figure 2 is an ideal signature. Difficulties of measurement happen both in Doppler characterisation and/or in phase detection when signals become distorted. This is notably so when the sensor interacts with very small drops or during tangential hits. Indeed, when the chord length becomes less than the sensitive length (the latter is 6 $\mu m$ here), partial wetting occurs and the signals exhibit limited amplitude as shown in Figure 3.

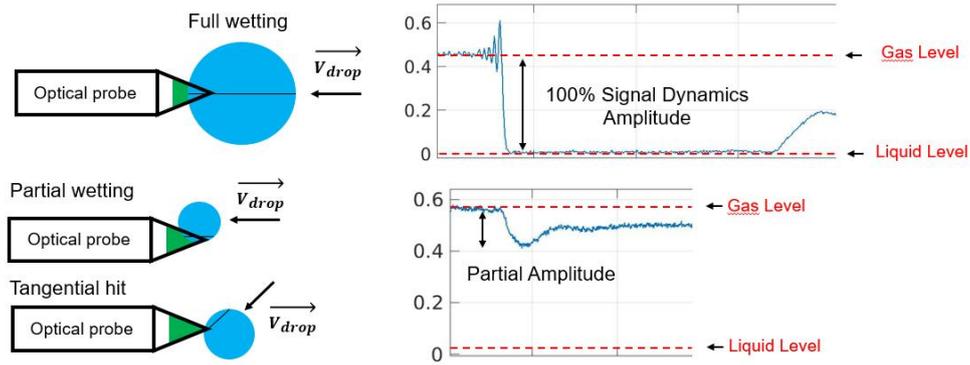

**Figure 3.** Illustration of the influence of the drop size and of the impact angle on signal amplitude. Optimal case of a drop with a size larger than or equal to the probe sensitive length, arriving aligned with the probe axis (top). Small drops touching the probe with a large eccentricity or angle (bottom). The dark lines inside drops are the chords measured.

When using phase detection probes in dense sprays, the gas level is no longer stable. The raw signals collected at $Q_G = 300$ and $2667\ L/min$ shown on **Figure 4** illustrate the strong differences occurring on the drop arrival frequency and on the gas level evolution with time. The $Q_G = 300\ L/min$ acquisition displays a low arrival frequency and a fairly stable gas amplitude. The high flow rate case exhibits many partial amplitude signals as well as an unstable gas level between signals. In such conditions, drop detection becomes very challenging and it requires a more refined processing routine. In particular, since the drop size decreases with the injected gas flow rate, the number of weak amplitude signals is expected to increase with $U_G$.

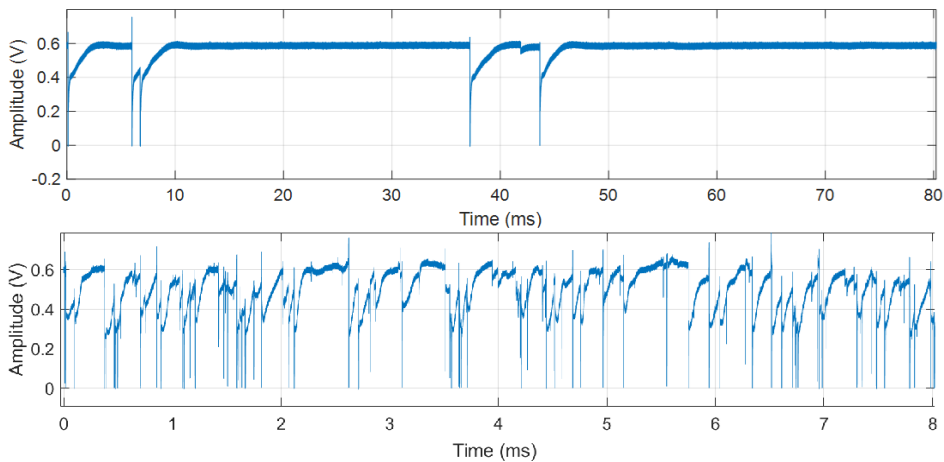

**Figure 4.** Samples of raw signals at low and high gas flow rates: $Q_G = 300\ L/min$ (top) and $Q_G = 2667\ L/min$ (bottom). $Q_L = 20\ L/h$. Note the difference in time scales on each plot. Probe was located 50 mm downstream the nozzle at the centre of the spray for $Q_L = 20\ L/h$.

Clearly, statistics on drop size and velocity as well as flux measurements would be biased if such distorted signals are not properly detected. We thus developed a processing able to detect all droplets over a wide range of injection conditions. Our objective was to quantify the amount of such distorted signals, and also to provide accurate drop size measurements. The developed processing we developed provides the amplitudes and the liquid residence time distributions. It also allows correlating the presence of Doppler signals with the signal amplitude with respect to phase detection.

A high pass Butterworth filter of order 2.1 MHz frequency is applied on the raw signal. All stable parts of the signal in amplitude then fall to 0 Volts (liquid plateau or stable gas level between two events for instance), only phase transitions remain non-zero. Note here that the high-pass cut off frequency depends on the steepness of phase transition gradient. Applying a negative voltage threshold, time locations of phase transitions can be detected. In a second step, one proceeds to local minima search starting from the phase transitions previously found. For each event *i* its amplitude is extracted (see Figure 5), and computed as:

$$Am(i) = V_{G,local}(i) - y_{min}(i) \tag{5}$$

where $V_{G,local}(i)$ is the local gas level preceding the event *i* and $y_{min}(i)$ the local minima. The liquid residence time $t_L(i)$ is then evaluated as the time spent by the signal in the liquid using a suitable threshold in amplitude. Finally, to correlate the presence of a Doppler burst with signal amplitude, we searched for the local maximim $y_{max}(i)$ occurring before the gas-liquid phase transition is found. If that maximum exceeds the local gas level $N(V_{G,local}(i))$, plus the noise at this level, a Doppler signature is assumed to be present ( (Figure 5). This processing enables to find all drop events those amplitude exceeds the maximum noise amplitude: this process is reliable but it involves long computation times.

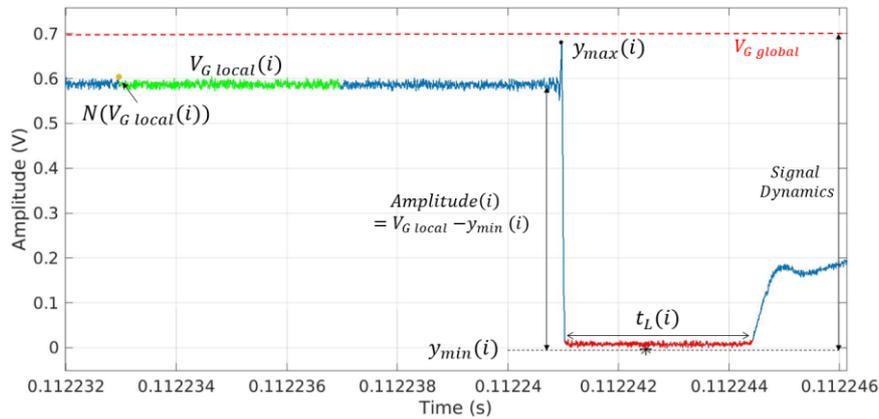

**Figure 5**. Illustration of the variables measured for each event *i*. The signal amplitude is the difference in voltage between the local minimum $y_{min}(i)$ and the local gas level $V_{G,local}(i)$ (highlighted in green). The liquid residence time $t_L$ is computed as the time interval around the local minima with a nearly constant amplitude (i.e. presenting a plateau, highlighted in red). $y_{max}(i)$ is the local maximum occuring before the phase transition: a Doppler signal is present if $y_{max}(i)$ exceeds the local gas level $V_{G,local}(i)$ plus the maximum noise amplitude.

**Results and Discussion**
### 1. Liquid phase detection

Experiments were run at four different gas flow rates $Q_G$ for a fixed liquid flow rate $Q_L$= 20 *L/h* (see Table 1). The probe was located on the axis of symmetry of the spray and 50 mm downstream the nozzle exit. For the conditions considered, that distance represents from twice the break-up length L$_B$ for U$_G$ = 32 m/s up to 3*L$_B$ at U$_G$=283 m/s.

The measurements were based on records 4 to 5 seconds long. Table 1 provides the liquid fractions and the mean drop arrival frequencies. Both quantities steadily increase with $Q_G$. In particular, the arrival frequency increases by a factor 1000 when the gas velocity increases by about a decade: this is because the drop size strongly diminishes with $U_G$.

**Table 1** – Injection conditions along with measured liquid fraction and mean drop arrival frequency ($f_{arrival}$).

| Run | $Q_L$ (L/h) | $U_L$ (m/s) | $Q_G$ (L/min) | $U_G$ (m/s) | Liquid fraction | Nb. of detected events | $f_{arrival}$ (Hz) |
|---|---|---|---|---|---|---|---|
| 1 | 20 | 0.28 | 300 | 32 | 0.0007 | 931 | $1.55 \times 10^2$ |
| 2 | 20 | 0.28 | 540 | 57 | 0.0023 | 3997 | $1.33 \times 10^3$ |
| 3 | 20 | 0.28 | 800 | 85 | 0.0026 | 14082 | $3.9 \times 10^4$ |
| 4 | 20 | 0.28 | 2667 | 283 | 0.0060 | 53214 | $7 \times 10^5$ |

The distributions of the signal amplitudes defined by Equation 5 are shown in Figure 6. The signal amplitudes are normalized by the signal dynamic SD defined here as the amplitude between the static gas level (i.e. the gas amplitude measured for a dry probe) and the minimum amplitude at the liquid level (see Figure 5). For all flow conditions, the signal amplitude recorded ranges from less than 0.1 up to 1. However, the shape of the distribution changes with the gas flow rate. At low gas flow rate, say below ≈600L/min, the fraction of full amplitude signals is quite large. That fraction decreases with $Q_G$ and, for the largest gas flow rate considered, it drops down to a ≈$10^{-2}$ probability. Conversely, partial amplitude signal becomes much more probable at high $Q_G$. This is the consequence of the lack of stability of the gas level exemplified in Figure 4. As the gas velocity increases, so does the drop arrival frequency. When the latter becomes too large, the probe tip never fully dries between successive drop impacts and that induces fluctuations in the amplitude of the signal corresponding to a probe tip presumably immersed in air. In addition, more drops and smaller drops mean more tangential impacts and thus a higher probability to partially wet the probe. Both effects lead to strong changes in the distribution of signal amplitude.

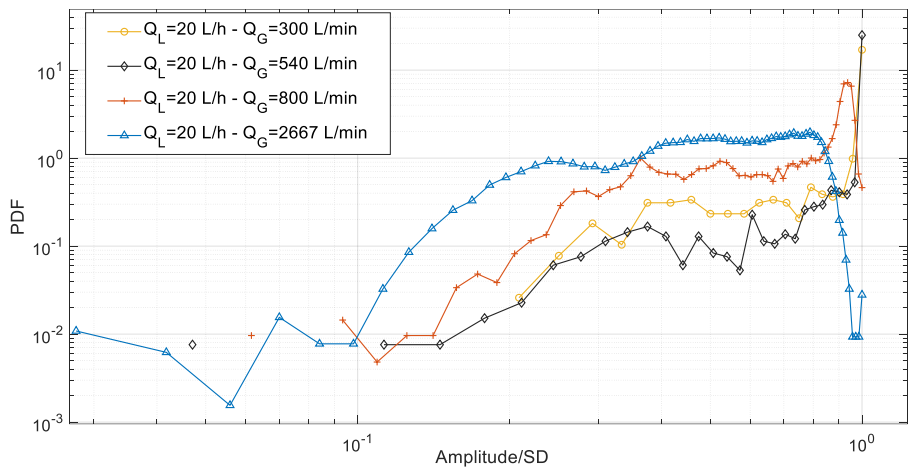

**Figure 6**. Distributions of signal amplitude appearing in the raw signal delivered by a Doppler probe in the experimental conditions given in **Table 1**.

The companion question is to determine which signals exhibit a Doppler burst. Figure 7 provides, for each class of normalized signal amplitude Am/SD, the percentage of signals for which a Doppler signal was detected. The main trend is that Doppler signals do exist for any signal amplitude between 0.05 and 1 SD. Beside, that result holds at any gas flow rate, except for the minimum observed at Am/SD=0.25 for $Q_G = 540\ L/min$ whose origin remains unknown. Although there are some fluctuations in Figure 7, the global trend is the same: there is a high (80 to 100%) probability to find a Doppler signal for amplitudes in the range 0.5 SD to 1.0 SD. Below 0.4-0.5 SD, the probability to find a Doppler signal can decrease down to about 55-60% depending on flow conditions. Consequently, all signal amplitudes can be exploited for velocity measurements, and the potential bias due to variations in signal amplitude remains limited.

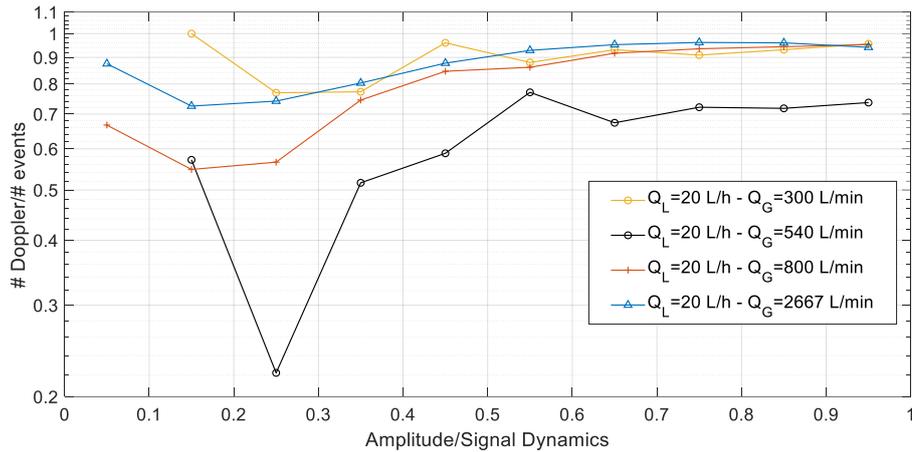

**Figure 7**. Percentage of signals possessing a Doppler burst according to each class in signal amplitude normalized by the maximum dynamics (bottom).

### 2. Drop velocity and size measurements using a Doppler processing routine

For the velocity measurements presented hereafter, and because of the very high sampling frequencies required (MHz at least), we used a simplified processing routine that only detected and analysed Doppler bursts on signals with a full amplitude plateau. In practice, only signals with an amplitude larger than 0.9 SD were considered. For data $Q_G = 300$ and $540\ L/min$ analysed in the previous section, these signals represent respectively 82 and 93% of the measured liquid fraction. Thus, an uncertainty of 10 to 20% on liquid time measurements has to be taken into account in what follows.

We first analysed the sensitivity of drop velocity measurements to the processing parameters used to select meaningful Doppler bursts. Satisfactory results were obtained when considering 10 successive oscillations and with a required temporal homogeneity better than 15%. These processing parameters have been kept the same. Again data were collected on the spray axis and at an axial distance z=50 mm from the injector exit.

As expected, the mean drop velocity monotonously increases with the gas velocity (Figure 8-right). The velocity distributions collected in the flow conditions of Table 1 are shown Figure 8-left. Interestingly, when centred, these distributions remain nearly the same whatever $U_G$. Small variations occur at the lowest velocities recorded (down to 0.2-0.3 the mean velocity) and also at the largest ones (3 to 5 times the mean). Note that the origin of these rare and very large velocities deserves to be investigated further. Indeed, with this sensor, a Doppler signal occurs only when the interface is within a few tens of micrometers from the fiber tip. Hence, and as we have shown in bubbly flows, such sensors may sometimes detect large velocities related with a very localized interface motion due for example to impacting droplets. In a first and crude attempt to eliminate these spurious events, we discarded 2/1000 of the largest velocities detected. The resulting drop size distributions are exemplified in Figure 9-left. The arithmetic mean chord $C_{10}$ is plotted in Figure 9-right together with the Sauter mean diameter evaluated as 3/2 $C_{10}$. As for velocity, the chord distributions measured for different gas flow rates almost collapse when they are centred. The distributions are quite merged around the mean, and the mean size decreases with the injected gas flow rate.

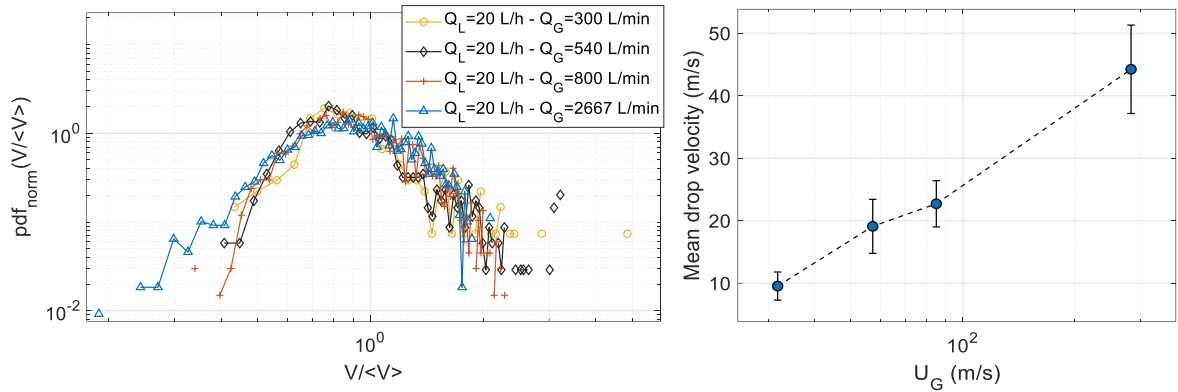

**Figure 8.** Evolution of the drop velocity distributions normalized by the mean (left) and of the mean drop velocity with gas velocity (right). Data collected on the spray axis and at an axial distance z=50 mm from the injector exit.

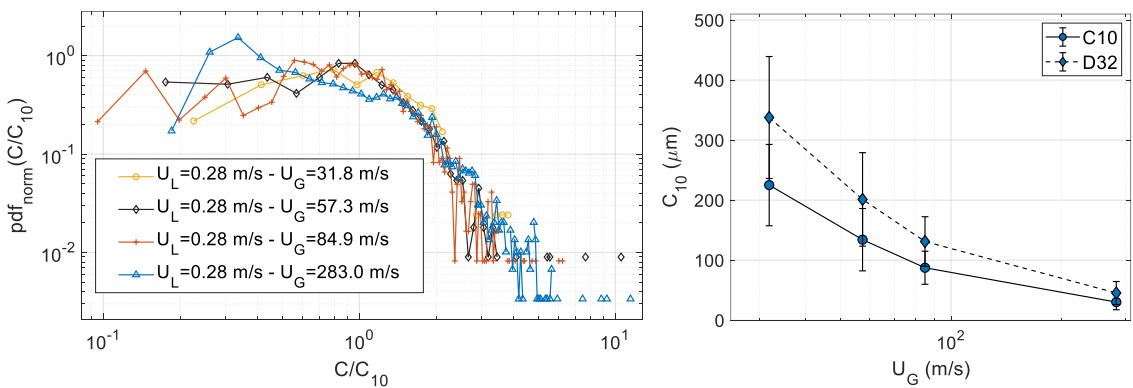

**Figure 9**. Droplet chord distributions normalized by the mean (left), mean chord and Sauter diameter evolution with the gas velocity at $U_L = 0.28\ m/s$ (right). For each set, 2/1000 chords with the greatest velocities have been deleted. Probe was located at the center of spray and at an axial distance z=50 mm from the injector exit.

### 3. Liquid flux measurements

Results of liquid flux measurements by using the simplified signal processing presented earlier for three liquid flow rates are presented in **Table 2**. Since the number of full amplitude signals prevails for flow rates up to $Q_G = 540\ L/min$ (cf. Figure 6), only flux measurements at fixed moderate gas flow are shown in this section. These fluxes were obtained with Equation 4. Since the signal processing routine is unable to detect small amplitude signals, in order to avoid a bias in the analysis, we considered a moderate gas flow rate, namely $Q_G = 460\ L/min$, for which the distribution in signal amplitude is rather uniform as shown on Figure 6. Radial profiles of the local liquid flux have been shown Figure 10. The difference between the flux recovered from local probe measurements and the global liquid flow rate is at most 8%. Since global flux measurements depend both on velocity and liquid residence time measurements, this proves the probe reliability under those injection conditions. These first results are quite encouraging: they deserve to be pursued on more complex flow conditions by combining the refined drop detection routine presented in section 1 with an efficient Doppler frequency detection.

**Table 2** – Liquid flux measurements for various liquid and gas flow rates. $Q_T$ is the flow rate computed from probe measurements over a spray diameter. $Q_{left}$ and $Q_{right}$ are the flux recovered on each half-spray section. *Probe location (x,y,z) [mm] varied on a spray diameter along x coordinate from $-Rmax$ to $+Rmax$ by steps of 1mm between each measurement. The axial distance is expressed both in absolute and relatively to $L_B$.

| $Q_L$ (L/h) | $U_L$ (m/s) | $Q_G$ (L/min) | $U_G$ (m/s) | $Q_{left}, Q_{right}$ (L/h) | $Q_T$ (L/h) | $Q_T/Q_L$ -- | Probe locations $(-R_{max}:R_{max}, y, z)$ |
|---|---|---|---|---|---|---|---|
| 10 | 0.14 | 460 | 48 | 5.1, 4.6 | 9.7 | 0.97 | (-11:11, 0, 30=2*$L_B$) |
| 20 | 0.28 | 460 | 48 | 9.7, 10.1 | 19.8 | 0.99 | (-15:15, 0, 30=1.5*$L_B$) |
| 30 | 0.42 | 460 | 48 | 13.7, 13.8 | 27.5 | 0.92 | (-16:16, 0, 38=1.5*$L_B$) |

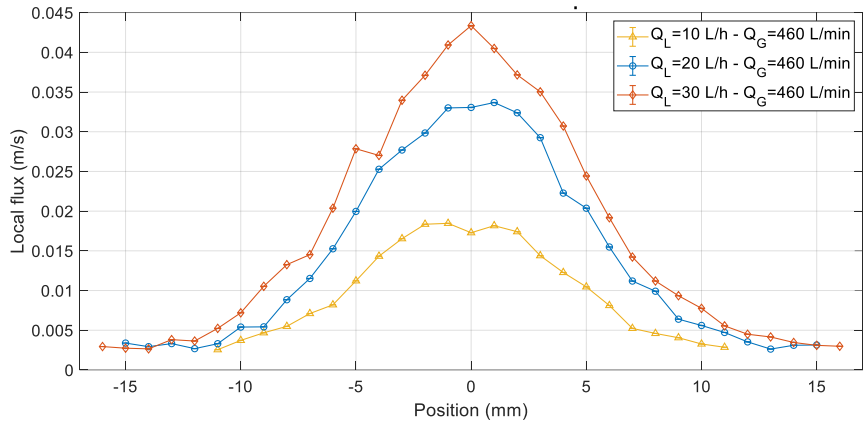

**Figure 10**. Radial profiles of the local liquid flux for different gas flow rates. Axial locations of the probe for each injection condition are provided in Table 2

## Conclusion

A new technique has been proposed for drop size and velocity measurements in sprays that combines an efficient phase detection by a conical optical probe with velocity measurements based on a Doppler shift on approaching gas-liquid interfaces. The analysis of raw signals collected in various conditions of assisted atomisation has shown that Doppler signals are indeed present even on signals of partial amplitude. Yet, as the spray gets denser, the probe tips remains wet and the signal delivered by the probe when its extremity is in the gas fluctuates a lot. That makes the phase detection routine quite complex.

In a first approach, we considered flow conditions such that these fluctuations remain limited and such that partial amplitude signals remain scarce. In these conditions, the spatial integration of local flux measured with the probe happens to be in very good agreement (≤8%) with the global liquid flow rate. These results are quite encouraging. They deserve to be complemented by a detailed investigation at larger gas flow rates where the raw signals from the Doppler probe become much more distorted.

## Nomenclature

| | |
|---|---|
| $Q_L$ | Injected liquid flow rate [L/h] |
| $Q_G$ | Injected gas flow rate [L/min] |
| $U_L$ | Injected liquid velocity [m/s] |
| $U_G$ | Injected gas velocity [m/s] |
| $V$ | Interface velocity [m/s] |
| $T_L$ | Liquid residence time [s] |
| $\alpha_L$ | Liquid fraction |
| $j_L$ | Local flux [m/s] |
| $V_{G\ global}$ | Global gas level [V] |
| $V_{G\ local}(i)$ | Local gas level of event i [V] |
| $y_{min}(i)$ | Minimum ordinate of event i [V] |